\documentclass[%
 aip,
 amsmath,amssymb,
 reprint,%
]{revtex4-1}

\usepackage{graphicx,color}
\usepackage{dcolumn}
\usepackage{bm}

\usepackage[utf8]{inputenc}
\usepackage[T1]{fontenc}
\usepackage{mathptmx}
\usepackage{etoolbox}
\usepackage{hyperref}

\newcommand{\beq}{\begin{equation}}
\newcommand{\eeq}{\end{equation}}
\newcommand{\bqa}{\begin{eqnarray}}
\newcommand{\eqa}{\end{eqnarray}}
\newcommand{\bal}{\begin{equation}\begin{aligned}}
\newcommand{\eal}{\end{aligned}\end{equation}}

\newcommand{\erf}[1]{Eq.~(\ref{#1})}

\newcommand{\ie}{{\it i.e.}}
\newcommand{\dg}{^\dagger}

\newcommand{\hs}{\ }

\newcommand{\balpha}{{\rm A}} 
\newcommand{\bbeta}{{\rm B}} 
\newcommand{\bgamma}{\Gamma} 
\newcommand{\click}{\star}

\newcommand{\erfand}[2]{Eqs.~(\ref{#1}) and (\ref{#2})}

\newcommand{\bra}[1]{ \langle{#1} |}
\newcommand{\ket}[1]{ |{#1} \rangle}

\newcommand{\ro}[1]{\left( {#1} \right)}

\newcommand{\red}{}

\newcommand{\blk}{}

\newcommand{\w}[1]{{#1}^{\rm w}}

\newcommand{\wv}[4]{{}_{\bra{#1}}\langle\w{#2}_{#3}\rangle_{\ket{#4}}}
\newcommand{\gwv}[4]{{}_{{#1}}\langle\w{#2}_{#3}\rangle_{\ket{#4}}}
\newcommand{\ti}{t_{\rm i}}
\newcommand{\tf}{t_{\rm f}}
\newcommand{\tm}{t_{\rm p}}

\newtheorem{theorem}{Theorem} 
\newtheorem{corollary}[theorem]{Corollary}
\newtheorem{lemma}[theorem]{Lemma}

\makeatletter
\def\@email#1#2{%
 \endgroup
 \patchcmd{\titleblock@produce}
  {\frontmatter@RRAPformat}
  {\frontmatter@RRAPformat{\produce@RRAP{*#1\href{mailto:#2}{#2}}}\frontmatter@RRAPformat}
  {}{}
}%
\makeatother
\begin{document}

\preprint{AIP/123-QED}

\title[Single-Photon Weak Values from Strong Coherent States]{Obtaining a  Single-Photon Weak Value from  Experiments using a Strong (Many-Photon) Coherent State}
\author{Howard M. Wiseman\vspace{1ex}}
\email{h.wiseman@griffith.edu.au, steinberg@physics.utoronto.ca, matin.hallaji@gmail.com}
\affiliation{Centre for Quantum Computation and Communication Technology (Australian Research Council), \\  Centre for Quantum Dynamics, Griffith University, Yuggera Country, Brisbane 4111, Australia\\ \vspace{1ex}}
\author{Aephraim M. Steinberg}%
\affiliation{Department of Physics, and Centre for Quantum Information and Quantum Control, \\ University of Toronto, 60 St.~George Street, Toronto, Ontario, Canada M5S 1A7 \\ \vspace{1ex}}%
 \affiliation{Canadian Institute For Advanced Research / Institut Canadien de Recherches Avancées, \\ 180 Dundas Street West, Toronto, Ontario, Canada, M5G 1Z8}
\author{Matin Hallaji}
\affiliation{Department of Physics, and Centre for Quantum Information and Quantum Control, \\ University of Toronto, 60 St.~George Street, Toronto, Ontario, Canada M5S 1A7 \\ \vspace{1ex}}%

\date{\today}

\begin{abstract}
A common type of weak-value experiment prepares a single particle in one state, weakly measures the occupation number of another state, and post-selects on finding the particle in a third state (a `click'). Most weak-value experiments have been done with photons, but the heralded preparation of a single photon is difficult and slow of rate. Here we show that the weak value mentioned above can be measured using strong (many-photon) coherent states, while still needing only a `click' detector such as an avalanche photodiode. One simply subtracts the no-click weak value from the click weak-value, and scales the answer by a simple function of the click probability. 
\end{abstract}

\maketitle

\section{\label{sec:level1}Introduction: Coherence, Productivity, Post-Selection, and Jon}


For the first and second authors, it is an alloyed pleasure to write a paper for this memorial volume celebrating Jonathan Dowling's academic  career. The passage of years gave us the distinction of being Jon's long-time sparring partners, and then took that away. We beg the reader's indulgence of our beginning the paper with some personal reminiscences, which also serve to say something about the topic of this paper. 
We will, eventually, motivate and present a new theory result at the intersection of  quantum optics and quantum measurements, where Jon did much of his work. Moreover, the result is simple to state, and relatively simple to derive, but not immediately obvious, so we think Jon would have liked it.

We have all appreciated Jon's iconoclasm. \red The first two authors (HMW and AMS) first encountered this during their PhDs, in Jon's work on classical and quantum effects in cavity QED. For AMS, this was through hearing Jon talk in person, an aural experience still many years away for HMW. For HMW, it was, instead, through the publication of Jon's delightfully titled paper on the topic,  ``Spontaneous emission in cavities: How much more classical can you get?''\cite{Dow93}. \blk That paper demonstrated that an effect then widely celebrated as a dramatic consequence of quantum mechanics could in fact be observed in a strikingly classical setting as well. As shall be seen, this is 
also a theme of the present work, in an entirely different context. 
\red In the years since, \blk both of us interacted with Jon many times, in many ways, some of which we will relate.

\begin{figure}
\includegraphics[width=0.9\columnwidth]{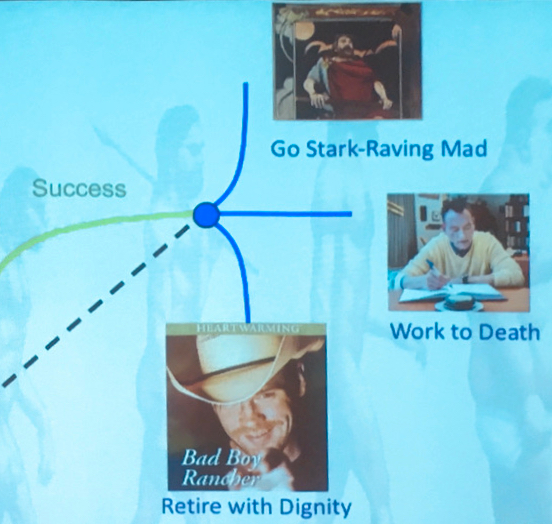}
\caption{\label{fig1} The Josephson Junction, as conceived by Jon Dowling. This is a photograph by the first author (HMW) of one of Jon's slides at the {\em Quantum Gates, Jumps, and Machines} conference, Brisbane, October 2018. For the significance of the top picture, see \href{https://howardwiseman.me/ThenArthurFought.html}{https://howardwiseman.me/ThenArthurFought.html}.}
\end{figure}

\begin{figure}
\includegraphics[width=1.0\columnwidth]{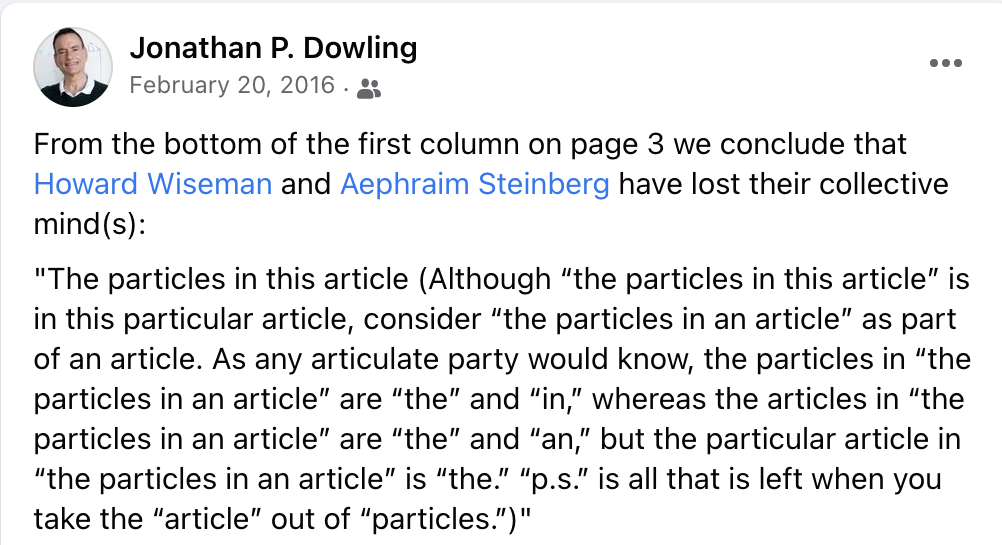} 
\caption{\label{fig1a} Further evidence provided by Jon in support of his case. Shockingly, said evidence was soon after removed by the editors of the (online only) journal in question\cite{Mah16}. 
}
\end{figure}

One way was that both HMW and AMS came to rely on Jon to do his utmost to help those of his colleagues more junior than himself. Another way relates to Jon's famous redefinition of the Josephson Junction as the ``time when a physicist must retire, go mad, or work to death,'' 
which the editors of this volume rightly celebrate. Jon, of course, chose the last path, all too literally. On one occasion, he seemed to suggest that HMW had chosen another; see Fig.~\ref{fig1}. (To be fair, he had further evidence, also implicating AMS; see Fig.~\ref{fig1a}.) It is certainly true that, unlike for HMW, Jon's rate of production only increased following that occasion, while suffering no loss of coherence. These qualities --- rate of production and coherence --- provide another weak connection to the scientific topic of this paper: the measurement of a single-photon weak value using multi-photon coherent states, which comes with an enormous advantage in rate of production.

\begin{figure}[b]
\includegraphics[width=1.0\columnwidth]{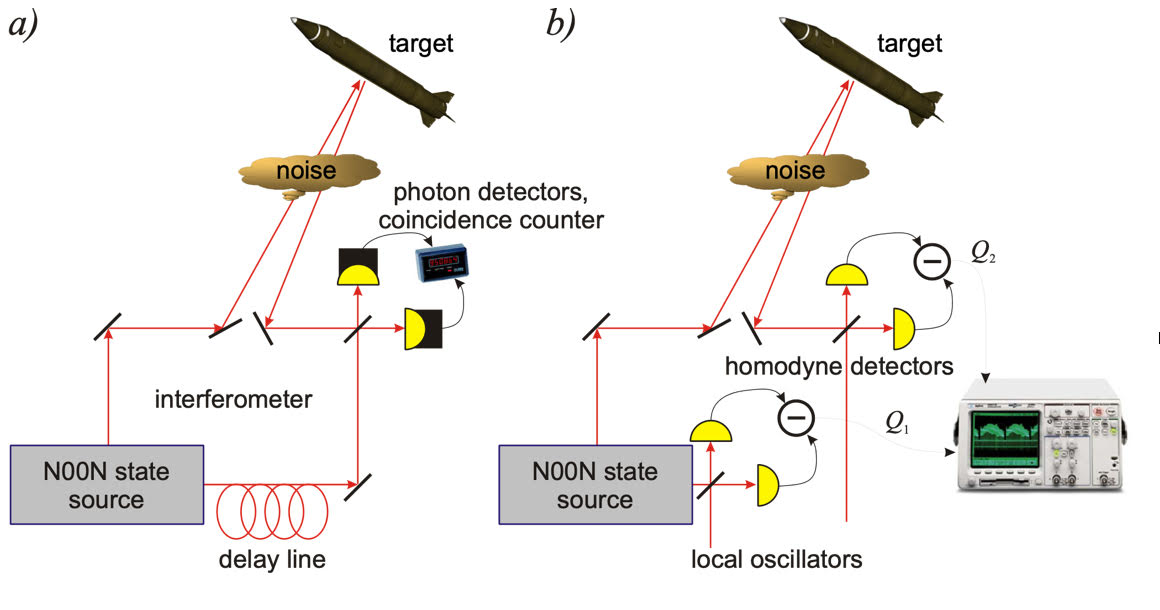} 
\caption{\label{fig1b}(a) Quantum imaging enables one to detect the perspicacity (target) behind Jon's seemingly insouciant iconoclasm (noise). (b) With post-selection.}
\end{figure}

A key component of weak values is post-selection\cite{AAV88}.   
Jon’s always creative vision made him an early adapter of photonic post-selection from 
linear-optics quantum computing to the task of preparing quantum states for quantum metrology or lithography\cite{LOLitho}.  This insight laid the groundwork for much of the work which the group of AMS went on to do on quantum metrology, and at one point led to Jon and AMS writing a grant proposal to pursue this further, along with Alex Lvovsky.  No matter how hard AMS tried, he could not dissuade Jon from including a picture of a missile in the optical schematic (see Fig.~\ref{fig1b}). Perhaps the fact that, in the end, Jon was funded, but instructed to jettison AMS and Alex, 
says something about the perspicacity behind his seemingly insouciant iconoclasm.  

\begin{figure}[t]
\includegraphics[width=0.8\columnwidth]{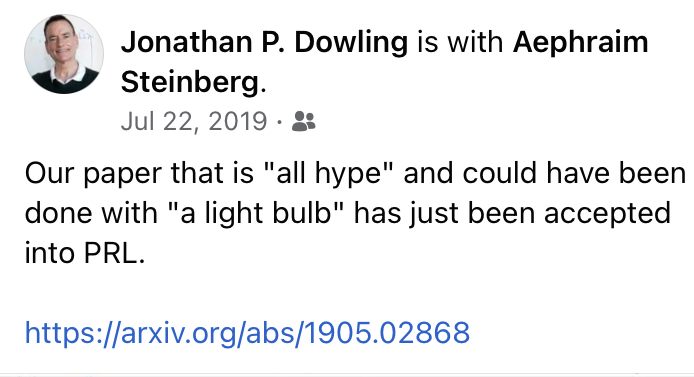} 
\caption{\label{fig1c} In another missile, Jon unfortunately misattributes\cite{Onanism} to AMS the claim that his PRL was ``all hype''. (AMS probably did say it could have been done with a light bulb.)}
\end{figure}

The present work applies similar ideas of post-selection, but in this case in order to draw conclusions about what single photons would have done from experiments performed using coherent states and avalanche photodiodes. One can only smile to imagine the pleasure Jon would have taken in mocking us for doing this, after having pretended to take offense at AMS's claim that a paper he was quite proud of\cite{lightbulb} could 
have been done “with a light bulb”; see Fig.~\ref{fig1c}.   

There are many reasons one might wish to study the evolution of individual quantum particles through the course of a process, for instance to ask how much we can say about ``where a photon has been'' as it traverses an interferometer\cite{UbiFuiste}, how much time a particle spends tunneling through a barrier\cite{ramos_spierings_racicot_steinberg_2020}, or whether or not a photon has been absorbed and re{\"e}mitted before being transmitted through a cloud of atoms\cite{sinclair-atom-photon-time}.  Yet while there have been dramatic advances in our abilities to produce and manipulate individual photons\cite{Wolfgramm,Steinlechner,Senellart}, the achievable rates are still far lower than the repetition rates of standard lasers.  It is therefore interesting to ask whether there are measurements one can do on coherent states which nevertheless reveal something about the behaviour of individual photons. \red It is well known that  G.I. Taylor's 1909 interference experiment\cite{Tay09}, using a very weak classical light source, was later \blk interpreted as confirming that ``a single photon interferes with itself.''  The quantum optics community long turned its back on such experiments, demanding nonclassical sources [{\em e.g.}, measurements that $g^{(2)}<1$]  in order to make statements about what individual photons actually do.  At the same time, many modern workers have fruitfully applied the idea that a sufficiently weak coherent state is essentially a superposition of the vacuum and the single-photon state, so that ``post-selecting'' on detection of a photon projects out only the latter term, and is hence guaranteed to produce exactly the same results as a true single-photon experiment (see for example\cite{Hulet,P1,Pittman2,Mir,Hosten,BenDixon,LB}).

While investigating the nonlinear phase shift which could be written by a single photon\cite{feizpour_hallaji_dmochowski_steinberg_2015}, we (AMS and co-workers)  found that the achievable shifts in our first experiment (${\cal O}(10 \mu {\rm rad})$) would be too small to measure practically with the heralded-resonant-photon rates we had at our disposal (${\cal O}(100 {\rm Hz})$).  On the other hand, it was straightforward to generate weak coherent states at much higher rates.  While at first we planned to limit these states to $|\alpha|^2 \ll 1$ so that only the vacuum and single-photon contributions were significant, we came to the realization that this restriction was unnecessary.  We showed that even if the experiment is done with a relatively strong coherent state at the input, in the situation where a detector placed at the output fires with low probability (e.g. because of collection or detection inefficiency), a click at that detector should lead one to update one's estimate of the average number of photons present in that trial by adding 1.  
Thus by subtracting the nonlinear phase shift observed when the detector did not fire from that observed when it did, we learned how large the single-photon phase shift was: although the experiment {\em always} had a ``background'' of 30-100 photons in the coherent state, each time the detector fired, there would be the additional effect of that one ``extra'' photon.  This led us to ask how far such reasoning could be pushed.  Could we talk sensibly about the effects, or the history, of ``that photon which had been detected''?  In particular, would an earlier proposal for ``amplifying'' single-photon nonlinearities via postselection\cite{Amir} work for photons post-selected in this fashion from a coherent source?  We (MH, AMS, and co-workers) went on to show that it could\cite{Matin}, in that specific case.

In this paper, we (the current authors)  give a much more general  proof that coherent states and click detectors can be used to measure single-particle occupation-number weak values, for arbitrary passive linear optical evolution before and after the weak measurement. The method is to subtract the no-click weak value from the click weak-value, and to scale this result by a simple function of the click probability. Thus, the click probability need not be kept small, and still  no calibration of the coherent intensity or detector efficiency is necessary. We begin in Sec.~\ref{sec:not} with notation and preliminaries, then the description of the experiment using coherent states in Sec.~\ref{sec:coh}. We give the main theorem, with proof, in Sec.~\ref{sec:thm}, before concluding with Sec.~\ref{sec:conc}.

\section{Weak Values: Notation and Preliminaries} \label{sec:not}

Consider the measurement of an observable $\Lambda$, represented by the Hermitian 
operator $\hat\Lambda$, using a weak coupling to a probe at time $\tm$. Let the coupling be minimally disturbing (in the sense defined in Ref.\cite{WisMil10}) and let the initial probe state be such that the variable that will carry the information about $\Lambda$ has Gaussian statistics. 
Say the system is prepared in state $\ket{\psi_{\rm i}}$ at time $\ti<\tm$ and is found by a projective measurement at time 
$\tf>\tm$ to be in state $\ket{\phi_{\rm f}}$. Denote the expectation value, over very many runs of an experiment, 
post-selected on the final measurement result, of the weak measurement result, as 
$\wv{\phi_{\rm f}(\tf)}{\Lambda}{\tm}{\psi_{\rm i}(\ti)}$. Let the system evolve unitarily apart from the measurements, 
as described by the unitary operator $\hat{U}(t',t)$ for evolution from $t$ to $t'$. 
Then, as shown by Aharonov, Albert, and Vaidman\cite{AAV88,AV90}, quantum mechanics predicts that, in the limit of arbitrarily weak coupling, this so-called {\em weak value} is arbitrarily well approximated by  
\beq \label{brawv}
\wv{\phi_{\rm f}(\tf)}{\Lambda}{\tm}{\psi_{\rm i}(\ti)} = \frac{\bra{\phi_{\rm f}}\,\hat U(\tf,\tm) \hat \Lambda \,\hat U(\tm,\ti) \ket{\psi_{\rm i}}}{\bra{\phi_{\rm f}} \,\hat U(\tf,\ti) \ket{\psi_{\rm i}}},
\eeq
provided the denominator is nonzero. 
Actually, as described above, the weak value will be the real part of \erf{brawv}. However, the imaginary part can also be found experimentally by measuring the property of the probe conjugate to that which carries the information about $\Lambda$. See a particularly careful recent experiment\cite{Vaidman2017-2}.  

Although this pure-state weak value has some unique features\cite{AV90,Vaidman2017-2}, it is useful to generalise this result\cite{Wis02}, and notation, to allow for 
an arbitrary final (post-selection) measurement, described by a positive operator $\hat E_{\rm f}$: 
\beq \label{effwv}
\gwv{E_{\rm f}(\tf)}{\Lambda}{\tm}{\psi_{\rm i}(\ti)} = \frac{\bra{\psi_{\rm i}} \,\hat U(\tf,\ti)\dg \hat{E}_{\rm f}\,\hat U(\tf,\tm) \hat \Lambda \,\hat U(\tm,\ti) \ket{\psi_{\rm i}}}{\bra{\psi_{\rm i}}\,\hat U(\tf,\ti)\dg \hat{E}_{\rm f}\,\hat U(\tf,\ti) \ket{\psi_{\rm i}}},
\eeq
In this regard, the following Lemma will be used later.
\begin{lemma} \label{lemma1}
Consider a bipartite ($A\otimes B$) system and let $\hat U(t',t)$ and $\ket{\psi_{\rm i}}$ be such that 
\beq \hat U(\tf,\ti) \ket{\psi_{\rm i}} = \ket{\psi^A_{\rm f}} \otimes\ket{\psi^B_{\rm f}},
\eeq
and consider $\hat E_{\rm f} = \hat{E}^A_{\rm f} \otimes \hat{I}$. Then 
\beq
\gwv{\left[{E}^A_{\rm f} \otimes {I}\right](\tf)}{\Lambda}{\tm}{\psi_{\rm i}(t_{\rm i})} = \gwv{\left[E^A_{\rm f}\otimes \ket{\psi^B_{\rm f}}\bra{\psi^B_{\rm f}}\right](\tf)}{\Lambda}{\tm}{\psi_{\rm i}(\ti)}.
\eeq
That is, the same weak value is obtained whether one post-selects the second subsystem to be in state $\ket{\psi^B_{\rm f}(\tf)}$ 
or one simply ignores it. 
\end{lemma}
{\em Proof.} Substituting both cases into \erf{effwv} gives the same expression, 
\beq
 \frac{\left( \bra{\psi^A_{\rm f}}\hat E_{\rm f}^A  \otimes\bra{\psi^B_{\rm f}} \right) \, \hat U(\tf,\tm) \hat \Lambda \,\hat U(\tm,\ti) \ket{\psi_{\rm i}}}{\bra{\psi^A_{\rm f}}\hat E_{\rm f}^A \ket{\psi^A_{\rm f}}}.
\eeq 
$\square$

\section{Weak Values with Coherent States} \label{sec:coh}
Consider a multipartite system in which each subsystem is a photonic mode. We will use the Greek letters $\alpha$, $\beta$, and $\gamma$ to denote 
coherent states. More specifically, we will use upper-case Greek letters ($\balpha, \bbeta, \bgamma$) for the full multimode system, lower-case Greek letters for one particular mode (the first), and lower-case Greek letters with an over-arrow for all the modes but the first. Thus, for example, 
\beq
\ket{\bgamma} \equiv \ket{\gamma,\vec\gamma} \equiv \ket{\gamma}\ket{\vec{\gamma}} \equiv \ket{\gamma}\ket{\gamma_1}\ket{\gamma_2} \cdots 
\eeq 
Say that $\hat U(t',t)$ is a passive linear optical transformation. That is, it takes a multimode coherent state to a multimode coherent state with the same mean photon number, $|\gamma|^2 + |\vec\gamma|^2$. Let 
the system begin in a coherent state $\ket{\psi_{\rm i}} = \ket{\balpha}$,  the weakly measured observable $\Lambda$ 
be the photon number $\hat n$ of some mode, and the post-selection be done by a photon number measurement on some final mode, yielding result $m \in \mathbb{N}_0$. 
Because $\hat U(t',t)$ allows the modes to be defined in any way that is convenient, without loss of generality we can take 
\beq 
\ket{\balpha} = \ket{\alpha,\vec{0}},\hs \hat{\Lambda} = \hat{n}\otimes \hat I := \hat b\dg \hat b \otimes \hat I ,\hs \hat E_{\rm f} = \ket{m}\bra{m} \otimes \hat I. 
\eeq
That is, the nominally first mode initially plays the role of the input pulse, then the weakly probed mode, then the mode on which the post-selection is made. This gives us the following short-hand notation: 
\beq
\wv{m}{n}{\tm}{\alpha}  := \gwv{\left[\ket{m}\bra{m} \otimes \hat I\right](\tf)}{[\hat n\otimes \hat I]}{\tm}{\alpha,\vec{0}\,(\ti)}.
\eeq

\begin{figure}
\includegraphics[width=0.9\columnwidth]{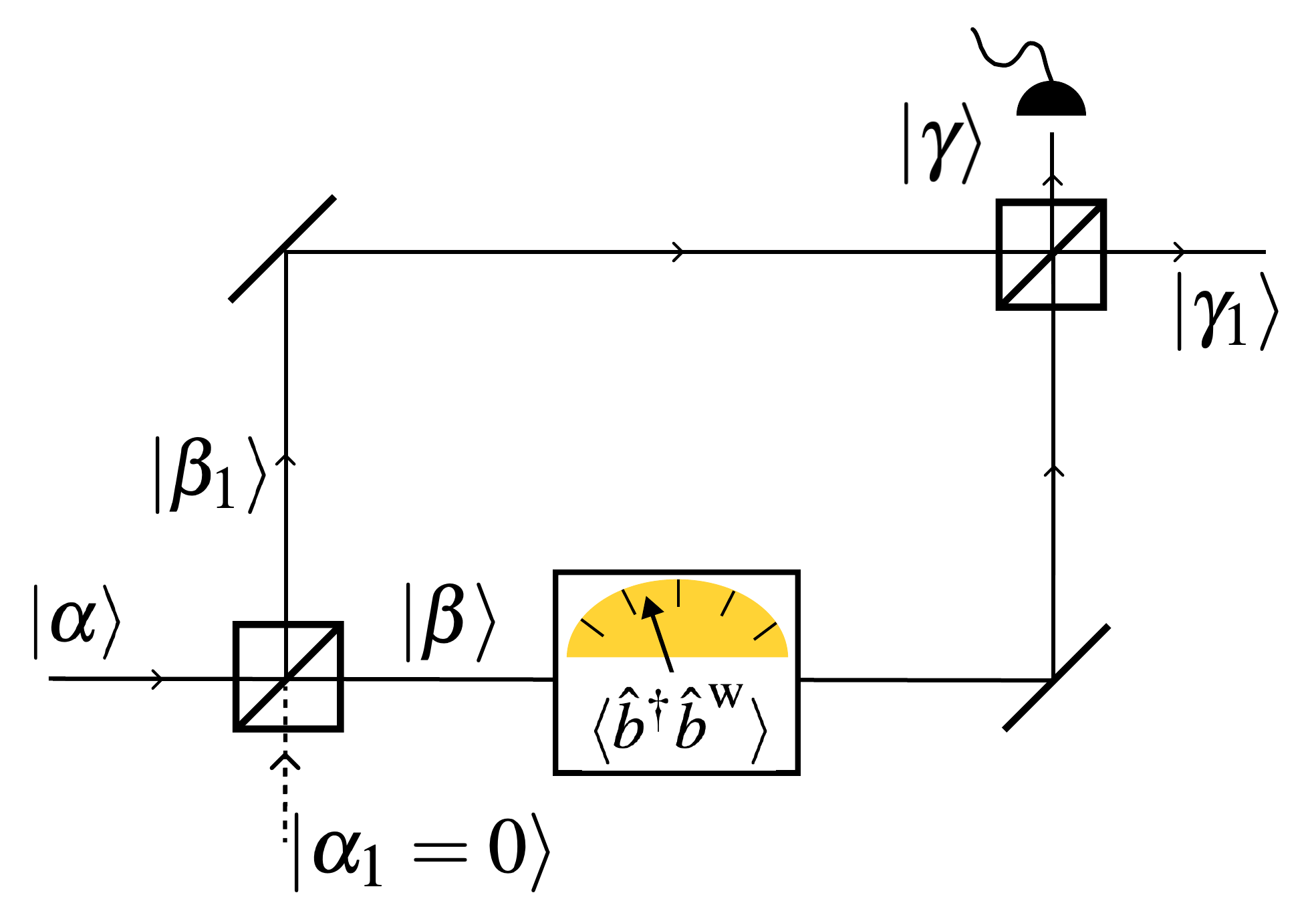}
\caption{\label{fig2} An example of the sort of situation under consideration, but with just two field modes in coherent states,  {\em i.e.}, $\ket{\alpha}\ket{\alpha_1}$, $\ket{\beta}\ket{\beta_1}$, and $\ket{\gamma}\ket{\gamma_1}$ at different stages in the evolution. Here the propagator 
$\hat U$ represents the evolution of the two modes through a Mach-Zehnder interferometer; a coherent state $\ket{\alpha}$ is incident from the left (at bottom), while vacuum is incident on the other port.  We are interested in a weak measurement of the photon number in the bottom path (that is, the operator $\hat b^\dagger \hat b$), after post-selection upon a click  at the detector in the upper port. The state entering that detector is, in the limit of arbitrarily weak measurement, arbitrarily well approximated by the coherent state $\ket{\gamma}$, where $|\gamma|^2=T|\alpha|^2$ as discussed in the text.}
\end{figure}

Let $\hat U(\tf,\ti) \ket{\balpha} = \ket{\bgamma}$.  Then from Lemma~\ref{lemma1}, we have 
\beq \label{fromlem}
\wv{m}{n}{\tm}{\alpha}  = \frac{\bra{m}\bra{\vec\gamma}\,\hat U(\tf,\tm) [\hat b\dg \hat b \otimes I] \,\hat U(\tm,\ti) \ket{\alpha,\vec{0}}}{\bra{m}\bra{\vec\gamma}\,\hat U(\tf,\tm) \,\hat U(\tm,\ti) \ket{\alpha,\vec{0}}},
\eeq
Now for a given $\hat U(\tf,\ti)$, there exists a $T$ (transmittance) such that, for all $\alpha$, 
\beq \label{defeta}
|\gamma|^2 = T |\alpha|^2 .
\eeq 

Next, we consider the special case $m=0$. This case is indeed special 
because it is the only post-selection case  where the final state is a coherent state. Specifically, defining
\beq
\ket{\bbeta} = \,\hat U(\tm,\ti) \ket{\alpha,\vec{0}}, \hs \ket{\tilde\bbeta_0} = \,\hat U(\tm,\tf) \ket{0,\vec{\gamma}}, 
\eeq
 it is easy to show from \erf{fromlem} that
\begin{align} 
\wv{0}{n}{\tm}{\alpha} = \tilde\beta_0^*\beta  \propto |\alpha|^2, \hs 
\gwv{}{n}{\tm}{\alpha} = |\beta|^2 \propto |\alpha|^2, \label{proptos}
\end{align}
where the (non-appearing) proportionality constants would be fixed for $\hat U(t',t)$ fixed. The second expectation value appearing in \erf{proptos}, the one without the left subscript, is the non-post-selected 
weak value. This of course coincides with the usual expectation value (for an arbitrary strength  measurement) given the initial state.

Now, most experiments with photonic weak values have been 
done with avalanche photo-diodes, or some other 
detector that does not resolve photon-number. (See Ref.\cite{Chantasri2021} for a recent survey of weak value theory and experiments.) 
Such detectors can still be sensitive to single photons, but by only giving two results: no `click', corresponding to $m=0$, and `click', corresponding to all $m>0$, described by the positive operator $\hat E_{\click} := \hat I - \ket{0}\bra{0}$. For short, we just use `$\click$' as the post-selecting subscript for a click, and from the operational definition of the weak value we have  
\beq 
\gwv{}{n}{\tm}{\alpha} = (1-p_{\click})\, \wv{0}{n}{\tm}{\alpha} \,+\, (p_{\click})\, \gwv{\click}{n}{\tm}{\alpha}, \label{alphasum}
\eeq
where $p_{\click}=1-\exp(-T|\alpha|^2)$ is the click probability. \red Equation~(\ref{alphasum}) can also be derived from \erf{effwv}, noting that $\hat I =  \ket{0}\bra{0} + \hat E_{\click}$. \blk One might worry  that the calculation here would be insufficient for describing inefficient detectors. However, this is not the case. Detector inefficiency can always be modelled by a beam-splitter 
before a perfect detector\cite{WisMil10}, 
 and this is included in the allowed passive transformation $\hat{U}(\tf,\ti)$. 
 \blk\blk

 \section{Weak Values with Single Photons, and the Theorem}\label{sec:thm}
 Now, consider the case where the input is a single photon,  $\ket{1}\ket{\vec{0}}$, 
 rather than a coherent state, but with the weak measurement and post-selection as before. In this case a click unambiguously means a single photon, so we denote the weak values by  $\wv{m}{n}{\tm}{1}$, where $m \in \{0,1\}$. Then, similar to \erf{alphasum}, 
 \beq 
 \gwv{}{n}{\tm}{1}  = (1-T) \wv{0}{n}{\tm}{1} + T \wv{1}{n}{\tm}{1}.\label{1sum}
\eeq
This follows since $T$ is the probability of a successful post-selection (`click') in the single-photon case. This can be seen from \erf{defeta} by considering the small $|\alpha|^2$ limit, to which we now turn. 
 
 There is a trivial and well-known (see in particular the Supplementary Information of Ref.\cite{feizpour_hallaji_dmochowski_steinberg_2015}) relation 
 of this single-photon weak value to the coherent-state one, in the limit $|\alpha|^2 \ll 1$. In this limit, the coherent-state input can be considered a 
 superposition of vacuum, with probability amplitude close to unity, and a single photon, with probability amplitude $\alpha$. Moreover, any of the multimode coherent states considered above, 
 such as $\ket{\bgamma}$,  can be considered a superposition of vacuum and a single photon split across those modes with probability amplitudes for the latter equal to the components of $\ket{\bgamma}$. Thus it is simple to see that 
 \begin{align} 
\gwv{}{n}{\tm}{\alpha}  &=  |\alpha|^{2}  \gwv{}{n}{\tm}{1} 
+ O(|\alpha|^4),  \label{alpha1corr}\\
 \gwv{\click}{n}{\tm}{\alpha}  &=  \wv{1}{n}{\tm}{1} + O(|\alpha|^2). \label{alpha1corr1}
\end{align}
The latter one follows because post-selecting on a click eliminates the vacuum component, and in the small $|\alpha|^2$ limit the single-photon term dominates. 

In contrast with the above relations --- which hold only in the limit of weak coherent states --- here we derive a non-trivial relation between the weak values in the single-photon and coherent state cases, which holds for arbitrary coherent states, as follows. 
\begin{theorem} 
For any finite $|\alpha|^2>0$, 
\beq \wv{1}{n}{\tm}{1} =  \left[\gwv{\click}{n}{\tm}{\alpha} -  \wv{0}{n}{\tm}{\alpha} \right] f(\red p_{\click}\blk),\label{thmeqn}
\eeq
where $f(x) := \red - \blk x/{\ln (1-x)}$. That is, the single-photon weak-value can be found 
from an experiment using a coherent state of arbitrary size, and an avalanche photodiode, by subtracting the no-click weak value from the click weak-value, 
and multiplying by a function $f\in\red(0,1)\blk$ of the click probability. 
\end{theorem} 
{\em Proof.}
 First, from the \red second \blk proportionality in \erf{proptos}, it follows that there are in fact no higher-order terms in \erf{alpha1corr}, and we can replace it by the exact expression
\beq \gwv{}{n}{\tm}{\alpha}  
=  |\alpha|^{2}  \gwv{}{n}{\tm}{1}. \label{alpha1corr2}
\eeq 
Second, we know from \erf{proptos} that 
$\wv{0}{n}{\tm}{\alpha} = O(|\alpha|^2)$ for small $|\alpha|^2$. Thus,  
expanding \erf{alphasum} in powers of $|\alpha|^2$ gives 
\beq 
\gwv{}{n}{\tm}{\alpha} = \wv{0}{n}{\tm}{\alpha} \,+\, T|\alpha|^2\, \gwv{\click}{n}{\tm}{\alpha} \,+\ O(|\alpha|^4). \label{smalphasum}
\eeq
Substituting into this \erfand{alpha1corr1}{alpha1corr2}  yields 
\beq 
 \wv{0}{n}{\tm}{\alpha} = |\alpha|^2 \left(  \gwv{}{n}{\tm}{1} - 
T\wv{1}{n}{\tm}{1}\right) + O(|\alpha|^4).
\eeq
Using \erf{1sum}, we obtain 
\beq 
 \wv{0}{n}{\tm}{\alpha} = |\alpha|^2(1-T)\wv{0}{n}{\tm}{1} + O(|\alpha|^4).
\eeq
But again using \erf{proptos}, it follows that the higher-order terms here are also zero, so we have another exact expression,  
\beq \label{alpha1corr3}
 \wv{0}{n}{\tm}{\alpha} = |\alpha|^2(1-T)\wv{0}{n}{\tm}{1}.
\eeq
Substituting \erfand{alpha1corr2}{alpha1corr3} into \erf{1sum} yields 
\beq
\wv{1}{n}{\tm}{1} = \ro{\gwv{}{n}{\tm}{\alpha} - \wv{0}{n}{\tm}{\alpha}}/T|\alpha|^2 . \eeq
Using (\ref{alphasum}) once more gives the desired result, (\ref{thmeqn}).
$\square$ \vspace{1ex}

\noindent {\em Remark}. 
It is no coincidence that the multiplicative factor $f(\red p_{\click}\blk)$ in \erf{thmeqn} is simply the reciprocal of the average number of photons  contributing to the click event ($m>0$). This average is $\bar{m}_{\click} \equiv \bar{m}/p_{\click}$, which evaluates to  $T|\alpha|^2/\left[1-\exp(-T|\alpha|^2)\right]$ 
for a Poisson distribution. In the limit of small $p_{\click}$, click events almost always involve only a single photon, and this factor goes to $1$. That is, in this limit, the single-photon weak value reduces to the difference between the click and no-click coherent weak values. 

\red 
\subsection{Dark Accounting}

As discussed already, detector inefficiency makes no difference to the above result. But there is another common detector imperfection: dark counts. Perhaps surprisingly, our result can very easily generalized to take this into account. Let $\check p$ be the probability of a dark count in the experiment if $\alpha=0$. For detectors with a finite dark-count rate, this would require defining a time-window for the experiment. Let us use $\check{0}$ and $\check\click$ to denote the events of a null-count and a non-null count in the situation where the non-null count could result from a dark count. Considering the probabilistic processes that can give rise to the two outcomes we are considering, we have
\begin{align}
p_{\check{0}} & = (1-p_\click)(1-\check{p}), \\ 
p_{\check\click} &= \check{p} + p_\click (1  -\check{p}).
\end{align}
Then, from the operational definitions of weak values, 
\begin{align}
\gwv{\check{0}}{n}{\tm}{\alpha} &= 
\wv{0}{n}{\tm}{\alpha},\\
p_{\check\click}\times \gwv{\check\click}{n}{\tm}{\alpha} &= 
p_\click \times \gwv{\click}{n}{\tm}{\alpha} + (p_{\check\click} -p_\click)\wv{0}{n}{\tm}{\alpha}.  
\end{align}
Rearranging, we obtain
\beq
p_{\check\click} \ro{ \gwv{\check\click}{n}{\tm}{\alpha} - \gwv{\check{0}}{n}{\tm}{\alpha} }
= p_\click \ro{ \gwv{\click}{n}{\tm}{\alpha} - \wv{0}{n}{\tm}{\alpha}}.
\eeq
From this, we derive a simple generalization of Theorem 2:
\begin{corollary} 
For any finite $|\alpha|^2>0$ and $\check{p}\geq 0$, 
\beq \wv{1}{n}{\tm}{1} =  \left[\gwv{\check\click}{n}{\tm}{\alpha} -  \gwv{\check 0}{n}{\tm}{\alpha} \right] 
g(\check{p},p_{\check\click}),\label{coroleqn}
\eeq
where (with $f$ as in Theorem 2), 
\beq 
g(y,z) := \frac{z(1-y)}{z-y}f\ro{\frac{z-y}{1-y}}
\eeq
That is, the single-photon weak-value can be found from an experiment using a coherent state of arbitrary size, and an avalanche photodiode with dark counts, by subtracting the no-click weak value from the click weak-value, 
and multiplying by a function $g\in(0,\infty)$ of the dark ($|\alpha|^2=0$) click probability $\check p$ and the non-dark (actual $|\alpha|^2$) click probability $p_{\check\star}$. 
\end{corollary} 

\blk

\section{Discussion} \label{sec:conc}

We have shown that weak-valued properties of a single photon can be determined experimentally using a coherent state input in place of a single-photon input, and a simple click/no-click detector.  Importantly, it is not necessary for the coherent state to be weak (\ie, have a mean photon number much less than one). 
This is of particular use in the case that the post-selection event in the single-photon case is rare, which is often the case for `interesting' weak value experiments, where an anomalous value (in this case, outside the range $[0,1]$) is sought\cite{Hulet,Pryde,LundeenHardy,BenDixon,Matin,Vaidman2017-2,Rebufello}. Thus, when using a coherent state, the click probability $p_\click$  could be made much larger than the single-photon click probability $T$, giving a much higher data rate than even heralded single photons. In addition, coherent states can, of course, be produced on demand much more easily than single-photon states. 

Another advantage of the coherent state option is that high-efficiency detectors are not needed. 
As discussed, detector inefficiency can always be modelled by a beam-splitter before the detection. From this it can be shown that a low-efficiency detector does not change the weak value in the single-photon case, but, because it reduces the click-probability $T$, it entails more runs being needed to obtain a large enough ensemble to give a weak value with the desired accuracy. In the coherent-state case, by contrast, the intensity $|\alpha|^2$ can simply be tuned to compensate for a low detection efficiency. \red Similar comments hold for optical losses in any part of the evolution, as long as the loss is the same along any path leading from the initial state to the detector. \blk

\red As we have shown, a minor modification of our protocol also allows the effect of detector dark counts to be removed. If dark counts were a significant problem, then even for a single-photon experiment one would have to subtract the no-click weak value from the click weak value to obtain the desired result, enfeebling any objection to the coherent-state method on the basis of its indirectness. 
An even simpler generalization of our result would be to allow more than one mode to be coupled weakly to the probe at time $\tm$, and to allow more than one mode to \blk have its photon number measured at time $\tf$. The proof is left as an exercise to the reader. 
On the other hand, it must be emphasized that our result holds only when the observable that couples to the probe in the experiment is the photon number of a mode (or set of modes, as just discussed). Without this, \erf{proptos} would not hold and the theorem would not go through. It is easy to construct a counter-example by considering the operator $\hat\Lambda = \hat I\otimes \hat I$.  

\red Given our peninitial paragraph, the reader may be wondering whether this penultimate paragraph will deliver the punch line ``Single-photon weak values: How much more classical can you get?'' Well, it just did, but in fact we argue the opposite. As mentioned above, the weak values we consider can be anomalous, such as when $\wv{1}{n}{\tm}{1}<0$. Moreover, Pusey\cite{Pus14} showed that any anomalous weak value is a proof of contextuality; such a thing cannot arise in a classical theory. 
Now, we have shown that $\wv{1}{n}{\tm}{1}$ can be obtained as the positively scaled {\em difference} of two differently post-selected averages of weak measurements of a positive operator ($\hat n$) with a coherent-state input, rather than as one such post-selected average with a single-photon input. Obviously a difference of two positive values can be negative, raising the possibility that in our scheme a classical explanation exists. However, our result includes as a special case the limit of a very weak input coherent state, in which case the subtracted value is negligible, meaning that the anomalous negativity of the weak value can arise from a single post-selected average with a coherent state. Therefore, despite the seeming classicality of coherent states and click detectors, there can be no classical explanation of the weak values in our general scenario. In this regard, one should remember that a non-destructive weak measurement of photon number is also an essential ingredient in the weak-value experiments we consider.

To conclude, \blk our result motivates further directions for investigation. We have considered photon detection which does not resolve among different positive photon numbers. 
As Remarked upon above, in the limit of small $p_\click$, the single-photon weak value reduces to the difference between the click and no-click weak values.  Moreover, for any value of $p_\click$, this difference is {\em exactly} equal to the single-photon weak value multiplied by the average number of photons involved in a ``click'' event.  This supports earlier approximate results which suggested that if a photon-number resolving detector were used, the weak value post-selected on detecting $m$ photons would exceed the no-click weak value by precisely $m$ times the single-photon number.  We leave a rigorous proof of this conjecture for future work.

\begin{acknowledgments}
This work was supported by the  Australian Research Council Centre of Excellence grant CE170100012, by NSERC (the Natural Sciences and Engineering Research Council of Canada), and the Fetzer Franklin Fund of the John E. Fetzer Memorial Trust.  We acknowledge the Yuggera people, the traditional owners of the land where this work was undertaken at Griffith University, and  the Huron-Wendat, the Seneca, and the Mississaugas of the Credit River, the traditional owners of the land on which the University of Toronto operates.  A.M. Steinberg is a Fellow of CIFAR.  F.M., P.M., and P.A.M. Steinberg all thank J.P. Dowling and D.R. Schmulian for having invented them and snuck them by countless editors over the years.  Jon and all of his alter egos are sorely missed.
\end{acknowledgments}

\subsection*{Data Availability Statement}
Data sharing is not applicable to this article as no new data were created or analyzed in this
study

\subsection*{Conflict of interest}
The authors have no conflicts to disclose.

\section*{REFERENCES}
\nocite{*}
\bibliography{WeakCoherent}

\begin{thebibliography}{32}
\bibitem [{Ona()}]{Onanism}%
  \BibitemOpen
  \href@noop {} {}\bibinfo {note} {In loving memory of Jon's refusal to ever
  let an editor have the last word, we were sorely tempted to accept our
  \LaTeX\ editor's suggestion to replace the apparently nonexistent word
  ``misattributes'' by (@hem) ``m@sturb@tes'', perhaps confirming another of
  his affirmations: that in Fig.~\ref{fig1a}. We were also tempted to place
  that figure here, as Jon would certainly have appreciated a footnote to a
  figure caption containing a captured figure quoting a figurative footnote,
  which was in itself equally self-referential and silly.}\BibitemShut {Stop}%
\end{thebibliography}

\end{document}